\documentclass[14pt
reprint,%
aps,%
groupedaddress,%
superscriptaddress,%
twocolumn,%
assume,%
amsmath,%
floatfix%,
]{revtex4-2}

\usepackage{graphicx}
\usepackage[
  colorlinks=true,
  urlcolor=blue,
  linkcolor=blue,
  citecolor=blue
]{hyperref}
\usepackage{siunitx,braket,amssymb,gensymb,ulem,color}

\newcommand{\rrm}[1]{\textrm{#1}}

\newcommand{\blue}[1]{\textcolor{blue}{#1}}

\newcommand{\dd}[2]{\frac{\rrm{d}#1}{\rrm{d}#2}}
\newcommand{\lr}[1]{\left(#1\right)}
\newcommand{\lrr}[1]{\left[#1\right]}

\newcommand{\apzt}{\alpha_{z\perp}}
\newcommand{\phzt}{\varphi_{z\perp}}
\newcommand{\apsp}{\alpha_{\rm spin}}
\newcommand{\phsp}{\varphi_{\rm spin}}
\newcommand{\apps}{\alpha_{\rm ps}}
\newcommand{\phps}{\varphi_{\rm ps}}
\newcommand{\sgz}{\sigma_z}
\usepackage{graphicx}% Include figure files
\usepackage{dcolumn}% Align table columns on decimal point
\usepackage{bm}% bold math
\usepackage{ORCIDinREVTeX}

\begin{document}

\preprint{APS/123-QED}

\title{Non-spreading meronic spin defects around optical vortices}

\author{Nilo Mata-Cervera}
\orcid{0000-0001-8464-5102}
\email{nilo001@e.ntu.edu.sg}
\affiliation{Centre for Disruptive Photonic Technologies, School of Physical and Mathematical Sciences, Nanyang Technological University, Singapore 637371, Republic of Singapore}
\affiliation{Complex Systems Group, ETSIME, Universidad Politécnica de Madrid, Ríos Rosas 21, 28003 Madrid, Spain}
\author{Miguel A. Porras}
\email{miguelangel.porras@upm.es}
\orcid{0000-0001-8058-9377}
\affiliation{Complex Systems Group, ETSIME, Universidad Politécnica de Madrid, Ríos Rosas 21, 28003 Madrid, Spain}
\author{Yijie Shen}
\orcid{0000-0002-6700-9902}
\email{yijie.shen@ntu.edu.sg}
\affiliation{Centre for Disruptive Photonic Technologies, School of Physical and Mathematical Sciences, Nanyang Technological University, Singapore 637371, Republic of Singapore}
\affiliation{School of Electrical and Electronic Engineering, Nanyang Technological University, Singapore 639798, Republic of Singapore}

\begin{abstract}
Optical vortices are singularity lines where the light field intensity vanishes and its phase is undefined. These threads of darkness are adorned by Gauss's law as lines of pure longitudinal polarization where the polarization plane tilts and winds around. The resulting spin field is as a unique structure with both features of topological texture and defect, as it includes a point defect of undefined spin enclosed by a meronic texture which spans half the spin unit sphere. This  topological structure does not spread in propagation: the normalized spin field maintains a deep-subwavelength confinement around the vortex line while the underlying vortex beam continues to diffract. 
% localized arbitrarily below the wavelength of light and presents highly anisotropic features. 
From their intrinsic nature and subwavelength confinement, these textures exhibit enhanced robustness upon isotropic perturbations such as atmospheric turbulence compared with other tailored polarization textures. Here we describe these topologies of transverse spin which decorate the phase singularity of paraxial vortex beams, highlighting the diversity of topological structures that arise in two different spaces---the spin unit sphere and the transverse-axial Poincar\'e sphere---and discuss the underlying aspects behind their subwavelength localization. 
\end{abstract}
\maketitle

\section{Introduction}
Optical singularities are loci at which the light field is not completely well defined~\cite{Berry2023ThePolarisation}. The interconnectedness of the diverse properties of light reveals singularities of intricate fields contained within singularities of more elementary ones~\cite{Vernon20233DFields,Shen2021SupertoroidalSpace}, just as phase singularities turn to be momentum or polarization singularities~\cite{Berry2009OpticalCurrents,Dennis2002PolarizationStatistics}. When a vector field vanishes at a certain point, this node is surrounded by a rich structure where the vector winds around, an arrangement that cannot be eliminated unless it annihilates with another one~\cite{Freund1995SaddlesFields}. Light fields have an extensive family of such singularities~\cite{Dennis2009SingularSingularities}, from points where the phase of vibration is undefined (optical vortices)~\cite{Coullet1989OpticalVortices}, to singularities of the polarization ellipse orientation~\cite{Berry2001PolarizationWaves}, energy flow~\cite{Wang2024TopologicalSkyrmions,Dow1967EnergyBeam} or spin angular momentum~\cite{Wang2022TopologicalLight}. 

\begin{figure}[b!]
    \centering
    \includegraphics[width=\linewidth]{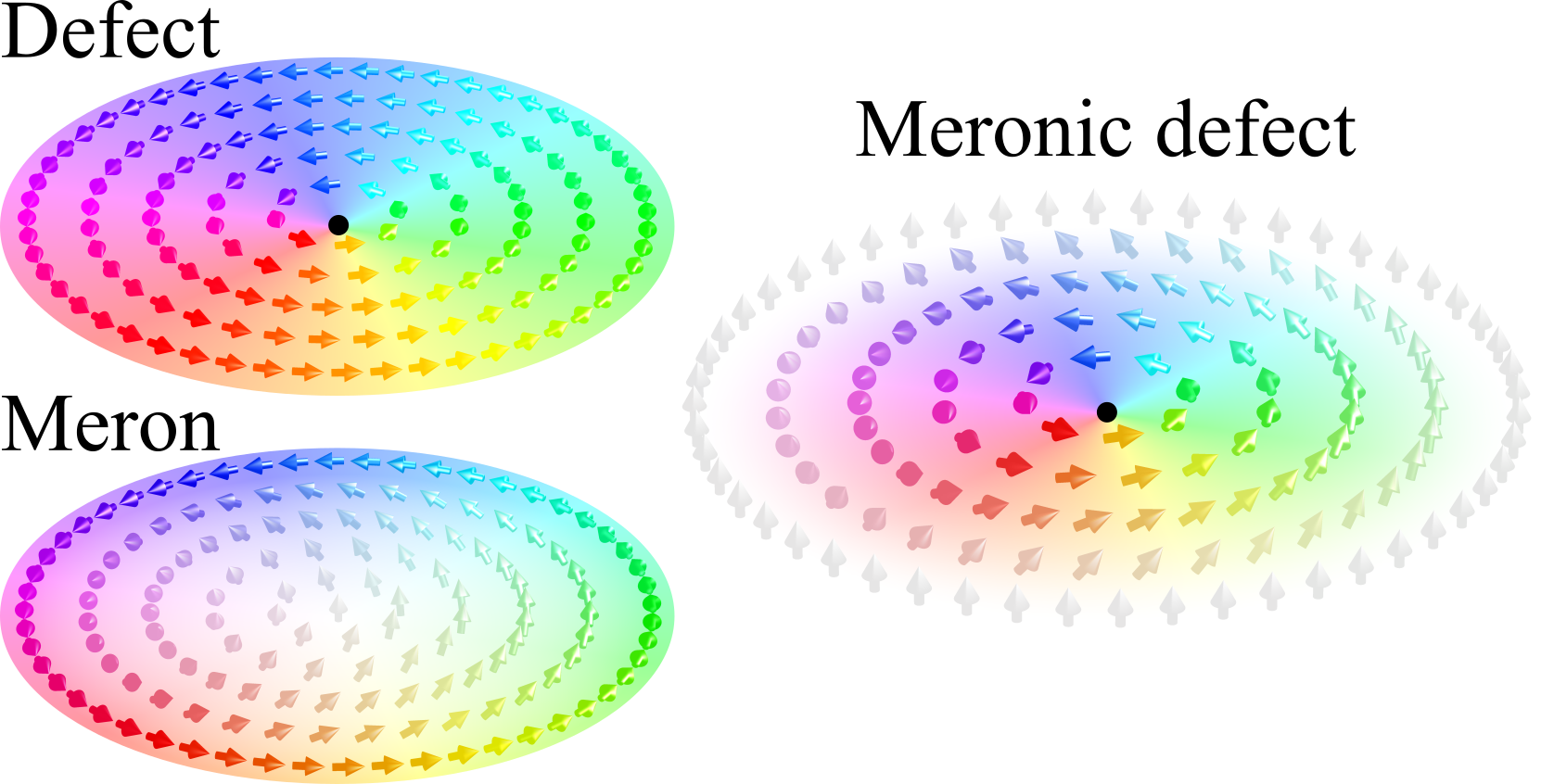}
    \caption{\textbf{Concept.} Point-defect in an azimuthally varying vector field (top left), meron texture in a 3D vector field spanning half of all possible orientations in the unit sphere (bottom left), and meronic defect combining both features (right).}
    \label{fig:concept}
\end{figure}

The intricate singular structure of highly confined light fields simplifies in the paraxial approximation, where light's polarization lays in the transverse plane, and linear and angular momenta point parallel to the propagation direction~\cite{Bliokh2015TransverseLight}. Yet, even simplistic arrangements such as a scalar vortex beam contain topological structures of non-transverse polarization and transverse spin. Around its phase singularity, where the longitudinal field dominates over the transverse one, the behavior of the field breaks with the most basic intuition about paraxial light waves: the polarization tilts from the transverse plane towards the propagation axis and rotates around it~\cite{Mata-Cervera2025SkyrmionicVortices}. This structure constitutes both a skyrmionic polarization texture and a meronic spin angular momentum defect (Fig.~\ref{fig:concept}) which arise naturally from Gauss's law in free-space. In contrast to all previous forms of polarization and spin textures, this polarization structure propagates without divergence attached to a vortex line, while the surrounding light field experiences standard diffraction~\cite{Afanasev2023NondiffractiveBeams,Vernon2024Non-diffractingRadiation,Mata-Cervera2026Diffraction-FreeVortices}. The geometry of these meronic spin defects is highly anisotropic, and is confined arbitrarily below the wavelength of light. 

\section{Texture-like defects}
Topological structures are usually classified into two categories~\cite{nakahara2018geometry,rajantie2002formation}:
\begin{itemize}
\item \textit{Topological defects}: These are special points, lines, and surfaces of discontinuous or undefined physical value, where the behavior of the singular properties around the discontinuity identifies topologically distinct field configurations (e.g. phase singularity at the center of a vortex beam, the spin defect at the center of a focused vector beam, and the L lines in a polarization vector field). 
\item \textit{Topological textures}: Patterns that continuously fill a surface or volume (in the absence of any singularity), where different topological arrangements are classified by the map between the physical space and order-parameter space (e.g. 2D meron, skyrmion, 3D hopfion, 4D skyrmion). 
\end{itemize}
The topological structure studied here exhibits both features of defect and texture simultaneously. Such structure includes a defect at the center as a singularity of vanishing spin, while in the vicinity of the defect the surrounding spin texture resembles a meron structure (mapping to a half 2-sphere). This contrasts with transverse spin vortices, where the vector field is contained in a plane and circulates around a nodal point of the field (whose orientation is fully undefined); and meron skyrmionic textures, where the vector field spans a half 2-sphere being well defined everywhere even if some components of the field vanish at specific locations (but the field itself is never zero). We denote such field as a meronic defect, see Fig.~\ref{fig:concept}. Hereinafter, we show how such unique meronic defects appear naturally around optical vortex lines in the vicinity of the phase singularity. 

Throughout this paper we describe textures of non-transverse polarization around the phase singularity of vortex beams. To show the orientation of any three-component vector $\bm{V}=\lr{V_1,V_2,V_3}$ we denote its normalized vector by lowercase $\bm{v}=\bm{V}/|\bm{V}|=\lr{v_1,v_2,v_3}$ (with $v_1^2+v_2^2+v_3^2=1$). We use spherical coordinates $\lr{\alpha,\varphi}$ where $\alpha=\cos^{-1}\lr{v_3}$ refers to the polar angle, the inclination angle with respect to the axis defined by $v_3$, and $\varphi=\tan^{-1}\lr{v_2/v_1}$ refers to the azimuthal angle, the rotation angle in the plane defined by $v_1$ and $v_2$. We use subscripts to distinguish spherical angles which refer to vectors in different parametric spaces: $\lr{\apps,\phps}$ for the Poincar\'e sphere, $\lr{\apzt,\phzt}$ for the transverse-axial Poincar\'e sphere and $\lr{\apsp,\phsp}$ for the spin angular momentum unit sphere, as defined later in sections~\ref{sec:2} and~\ref{sec:3}.

\section{Non-transverse polarization around vortex lines}\label{sec:2}
Consider the most elementary type of vortex beam: uniform polarization and cylindrically symmetric intensity. The intensity vanishes at the center, the phase singularity. Such wave field itself satisfies Helmholtz equation, or its paraxial counterpart, and as any type of structured wave it is naturally accompanied by a faint longitudinal field, making the entire electric field consistent with Gauss's law in free space $(\nabla\cdot\bm{E}=0)$. While the intensity of the axial field is generally insignificant in paraxial waves, it plays a predominant role at regions where the transverse field vanishes, near its phase singularities. As a result, the polarization state inevitably becomes structured around the zeros of a uniformly polarized transverse field. Coupling of field components through Gauss's law is reminiscent of the vector nature of light, evoking that even in simple settings light cannot be treated as a scalar wave. The complex envelope $\bm{\psi}$ of the electric field $\bm{E}=\bm{\psi}e^{ikz}$ reads
\begin{equation}\label{eq:psi}
    \bm{\psi}=\psi_\perp \bm{u}_\perp+\psi_z\bm{u}_z=\psi_\perp\lr{\bm{u}_\perp+\rho e^{i\gamma}\bm{u}_z},
\end{equation}
where the longitudinal field $\psi_z=\psi_\perp\rho e^{i\gamma}$ is expressed with respect to the transverse field through the field ratio $\rho e^{i\gamma}=\psi_z/\psi_\perp$, $k=2\pi/\lambda$ is the wavenumber for a monochromatic field with wavelength $\lambda$. The uniform transverse polarization $\bm{u}_\perp$ can be written as
\begin{equation}\label{eq:u_perp}
    \bm{u}_\perp=\cos\lr{\apps/2}e^{-i\phps/2}\bm{u}_R+\sin\lr{\apps/2}e^{i\phps/2}\bm{u}_L
\end{equation}
where $\apps$ and $\phps$ are the polar and azimuthal angles in the Poincar\'e sphere (PS), the parametric space representing all transverse polarization states~\cite{Alonso2023GeometricTutorial}, $\bm{u}_{L,R}=(\bm{u}_x\pm i\bm{u}_y)/\sqrt{2}$ are the Jones vectors of left- and right-handed circular polarization (LCP and RCP respectively). The orientation of the polarization ellipse in the transverse plane is given by $-\phps/2$ and its ellipticity angle $\chi=\apps/2-\pi/4$. RCP and LCP correspond to north $\lr{\apps=0}$ and south $\lr{\apps=\pi}$ poles of the PS respectively, where the azimuth $\phps$ is singular and so is the orientation of the polarization ellipse. The relative weight between RCP and LCP determines the expected value of the $z-$component spin angular momentum $\sgz$~\cite{Barnett2001OpticalAngular-momentumflux}, which for uniform transverse polarization (\ref{eq:u_perp}) is related to the third of the Stokes parameters $\sgz=-\cos\lr{\apps}$, yielding $\sgz=+1$ for LCP, and $\sgz=-1$ for RCP. Ignoring global constants and envelopes, the transverse complex amplitude around the phase singularity can be written as
\begin{equation}\label{eq:psi_perp}
    \psi_\perp(r,\phi,z)\propto\lr{\frac{r}{w(z)}}^{|\ell|}e^{i\ell\phi},
\end{equation}
where $w(z)$ is a $z-$dependent scaling parameter which accounts for the divergence of the beam, $r=\sqrt{x^2+y^2}$ and $\phi=\tan^{-1}\lr{y/x}$ are the radial and azimuthal coordinates in the transverse plane~\cite{Dennis2009SingularSingularities}. Once the transverse field is known, the longitudinal field is fixed by Gauss's law, which for paraxial waves reads $\psi_z=(i/k)\nabla_\perp\cdot\bm{\psi}_\perp$, where $\nabla_\perp\cdot$ is the transverse divergence~\cite{Lax1975FromOptics}. The longitudinal field associated with the vortex beam in Eq. (\ref{eq:psi_perp}) dominates around its phase singularity for $\ell>0$ and RCP, and for $\ell<0$ and LCP. On the other hand, it becomes negligible with respect to the transverse field around its phase singularity for $\ell<0$ and RCP, and for $\ell>0$ and LCP~\cite{Mata-Cervera2026Diffraction-FreeVortices}. For elliptical polarization the longitudinal field can be reconstructed from these two cases yielding
$$\psi_z=\frac{i|\ell|}{kr}\lr{\frac{r}{w(z)}}^{|\ell|}\sqrt{1-\rrm{sgn}\lr{\ell}\sgz}$$
\begin{equation}\label{eq:psi_z}
    \times \exp\lrr{i\rrm{sgn}\lr{\ell} \lr{|\ell|-1}\lr{\phi+\phps/2}},
\end{equation}
expressed in terms of the spin charge $\sgz$ and the sign of the topological charge $\rrm{sgn}\lr{\ell}$. Around the phase singularity, a non-transverse polarization structure emerges if the field ratio between transverse and longitudinal components is non-trivial. From (\ref{eq:u_perp}) and (\ref{eq:psi_z}) we obtain
\begin{equation}\label{eq:ratio}
    \rho e^{i\gamma}=\frac{\psi_z}{\psi_\perp}=\frac{i|\ell|}{kr}e^{-i\rrm{sgn}\lr{\ell} \lr{\phi+\phps/2}}\sqrt{1-\rrm{sgn}\lr{\ell}\sgz}.
\end{equation}
If the longitudinal field is negligible the ratio becomes trivial $\psi_z/\psi_\perp=0$ and the non-transverse polarization texture is lost. We remark that the polarization structure around the vortex core given by Eq. (\ref{eq:ratio}) is indeed independent of the shape of the vortex beam, as shown in the Appendix (\ref{sec:A5}). Yet, our analysis is limited to paraxial fields, therefore we neglect spin-orbit coupling and other depolarization effects arising in tight-focusing conditions~\cite{Gu2000AdvancedTheory}. 

Eq. (\ref{eq:ratio}) is analogous to the quotient between $x-$ and $y-$polarized fields introduced by Poincar\'e to describe the transverse polarization~\cite{Poincare1889TheoriePremier}, but here it involves the transverse and longitudinal complex amplitudes of the wave. The resulting transverse-axial (TA) polarization state can be represented parametrically by expressing (\ref{eq:ratio}) in the Riemann sphere~\cite{note1}. In this parametric space north and south poles are points of longitudinal and transverse polarization. The degree of longitudinal polarization is given by the polar angle 
\begin{equation}\label{eq:apzt}
    \apzt=\cos^{-1}\lr{\frac{|\psi_z|^2-|\psi_\perp|^2}{|\psi_z|^2+|\psi_\perp|^2}}=\cos^{-1}\lr{\frac{\rho^2-1}{\rho^2+1}}, 
\end{equation}
while the azimuthal angle represents the phase difference
\begin{equation}\label{eq:phzt}
    \phzt=-\arg\lr{\psi_z\psi_\perp^{*}}=-\gamma.
\end{equation}
We refer to this space as the transverse-axial Poincar\'e sphere (TA-PS) characterized by the spherical angles $\apzt$ and $\phzt$ or by the TA-Stokes vector $\lr{\sin\apzt\cos\phzt,\sin\apzt\sin\phzt,\cos\apzt}$.

The spatially-varying field ratio (\ref{eq:ratio}) in fact defines a skyrmionic polarization texture, a topological map from the TA-PS to the transverse plane~\cite{Mata-Cervera2025SkyrmionicVortices}. The polarization texture covers all the surface of this parametric sphere, from the north pole $\apzt\to0$ which is mapped to the vortex core $\lr{\rho\to\infty}$, gradually descending to the south pole $\apzt\to\pi$ as the polarization becomes transverse $\lr{\rho\to0}$ away from the phase singularity. All azimuths on the TA-PS $\phzt=-\gamma$ are present as we complete a full turn around the phase singularity. Thus, polar and azimuthal angles in the TA-PS are mapped into radial and azimuthal coordinates of the transverse plane, $\apzt=\apzt(r)$ and $\phzt=\phzt(\phi)$ respectively, a mapping with nonzero skyrmion number~\cite{Mata-Cervera2025SkyrmionicVortices}
\begin{equation}\label{eq:Qsk_zt}
    Q_{z\perp}^\star=\iint \sin\apzt\dd{\apzt}{r}\dd{\phzt}{\phi}\rrm{d}r\rrm{d}\phi=\rrm{sgn}\lr{\ell}.
\end{equation}
This skyrmion number is only valid for the unstable configuration (\ref{eq:psi_perp}) and is bounded to $\pm1$, we denote this case with the superscript ($^\star$). In real settings the skyrmion number $Q_{z\perp}$ becomes unbounded when the phase singularities of (\ref{eq:psi_perp}) are unfolded, as we discuss later in~\ref{sec:TC}.

Unlike in a paraxial skyrmionic beam where the transverse polarization ellipse features all ellipticities and in-plane orientations, the TA skyrmionic texture (\ref{eq:ratio}) is only parametric. The TA-PS captures information of the relative phase and amplitude between transverse and longitudinal fields, but does not provide a direct connection with the geometry of the 3D polarization ellipse. The later depends on the transverse polarization itself (\ref{eq:u_perp}), which is absent in the construction of the TA-PS (\ref{eq:apzt},~\ref{eq:phzt}). Yet, the polarization texture (\ref{eq:ratio}) can be described using physical magnitudes such as the 3D orientation of the polarization plane. As we discuss in the next section, when we express the features of the polarization texture (\ref{eq:ratio}) in physical space, the emerging topologies are richer and differ with respect to the highly symmetric skyrmionic textures in the TA-PS (\ref{eq:apzt},~\ref{eq:phzt},~\ref{eq:Qsk_zt}). 

\section{Meronic spin defects}\label{sec:3}
Non-transverse polarization is connected to transverse spin angular momentum (SAM) density, emerging at points where the polarization ellipse is not orthogonal to the wave's mean linear momentum~\cite{Bliokh2015TransverseLight}. For simplicity, we are only interested in the electric polarization ellipse, therefore ignoring constants the SAM can be evaluated from (\ref{eq:psi}) as $\bm{S}=\rrm{Im}\lr{\bm{\psi}^{*}\times\bm{\psi}}$~\cite{Berry2009OpticalCurrents,Vernon2024ADensity}. 
The three-component SAM density can be expressed in terms of the TA field ratio (\ref{eq:ratio}), the full expression can be found in Eq. (\ref{eq:S_xyz}) of~\ref{sec:A1}. We only focus on the direction of the SAM (unit vector orthogonal to the polarization ellipse), expressed in terms of azimuthal $\phsp=\tan^{-1}\lr{S_y/S_x}$ and polar $\apps=\cos^{-1}\lr{S_z/|\bm{S}|}=\cos^{-1}\lr{s_z}$ angles
\begin{eqnarray}
    \phsp=-\frac{\phps}{2}    -\tan^{-1}\lr{\frac{\tan\lr{\phzt}}{\tan\lr{\chi}}},\label{eq:phsp}\\
    s_z=\cos\lr{\apsp}=\frac{\sgz}{\sqrt{\sgz^2+2\rho^2\lr{1-\cos\lr{2\gamma}\sqrt{1-\sgz^2}}}}. \label{eq:sz}
\end{eqnarray}
Eq. (\ref{eq:phsp}) connects the azimuthal angle of the SAM density (measured relative to the transverse polarization orientation $-\phps/2$) with the azimuth of the TA-PS $\phzt=-\gamma$, the ellipticity angle of the transverse polarization $\chi$ scaling their relationship. From (\ref{eq:phsp}) we identify the poles of the TA-PS (singular $\phzt$) as singularities of the transverse SAM (singular $\phsp$), points of undefined azimuthal orientation of the polarization plane. These are also the phase singularities of the transverse and longitudinal fields (where $\gamma$ is singular). Intuitively, at the north pole of the TA-PS the polarization is longitudinal and thus the polarization plane is not defined (the ellipse degenerates to an axial line) and at the south pole the polarization plane lays completely on the transverse plane thus its azimuth is also undefined. Eq. (\ref{eq:phsp}) and (\ref{eq:sz}) provide a representation of the TA polarization structure (\ref{eq:ratio}) using the real space polarization plane orientation, in contrast to the coordinates in the TA-PS or the TA-Stokes vector which themselves only refer to an abstract parametric space. The resulting SAM texture constitutes a meronic defect, a spatially-varying SAM field with a singularity $\bm{S}=0$ at the centre. 
\subsection{Isotropic textures}
\begin{figure}[t!]
    \centering
    \includegraphics[width=\linewidth]{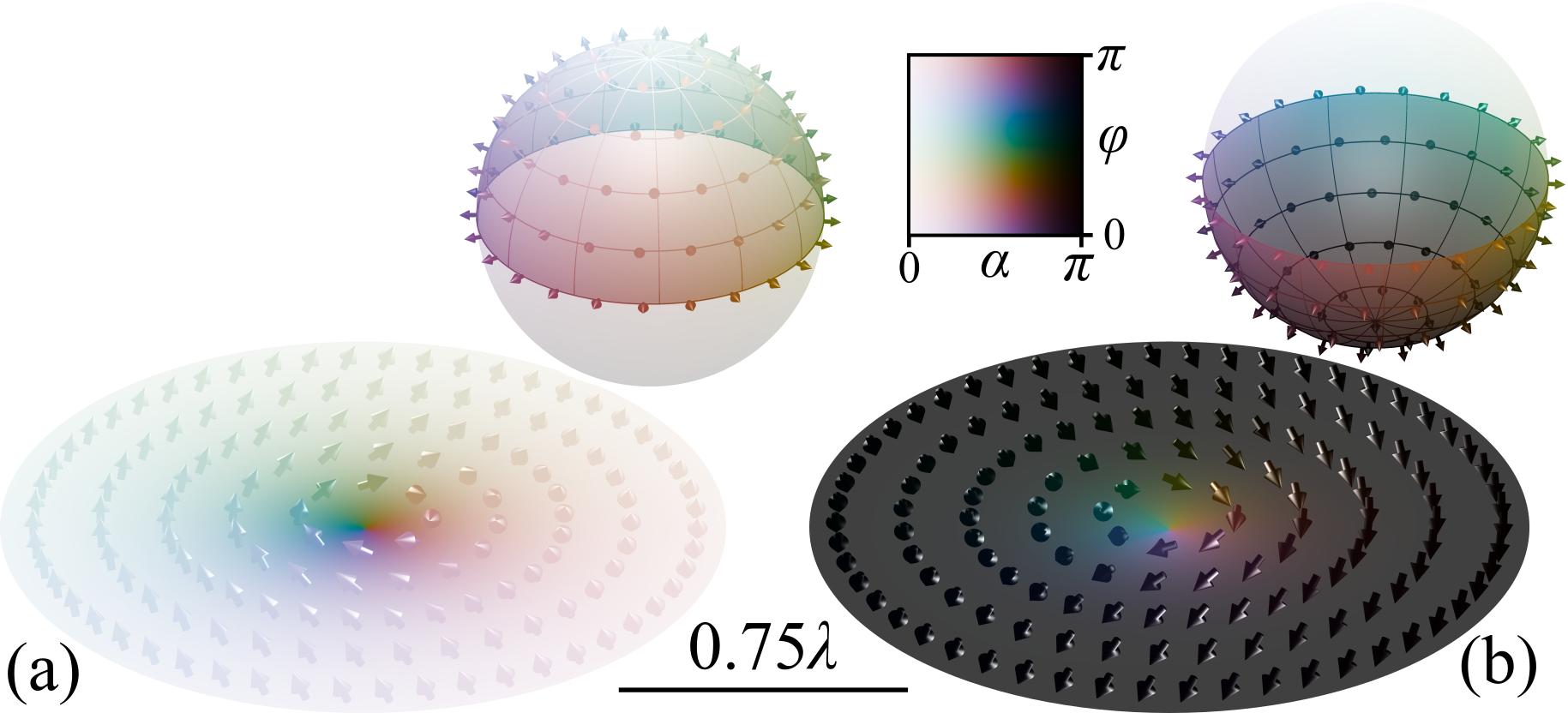}
    \caption{\textbf{Meronic spin defects around vortices.} SAM texture around the phase singularity of a vortex beam with $\ell=-1$ and LCP (a) and the same for $\ell=+1$ and RCP (b). The area covered in the SAM unit sphere is highlighted as an inset. Hue and lightness represent the azimuthal and polar angles in the unit sphere (inset).}
    \label{fig:F1}
\end{figure}

The simplest SAM texture (\ref{eq:phsp}, \ref{eq:sz}) appears when the transverse polarization is circular $\sgz=\pm1$, in this case the SAM azimuth simply becomes the TA-PS azimuth, and is the particular case reported in~\cite{Mata-Cervera2025SkyrmionicVortices}. The orientation $\phps$ in (\ref{eq:phsp}) is undefined for circular polarization but it factors out (see \ref{sec:A2}) and the azimuthal and polar angles of the SAM become only functions of the azimuth and radial coordinates in the transverse plane
\begin{eqnarray}
    \phsp(\phi)=\phi-\frac{\pi}{2}, \label{eq:phspcirc}\\
    s_z(r)=\cos\lr{\apsp}=\frac{\rrm{sgn}\lr{\sgz}}{\sqrt{1+4|\ell|^2/k^2r^2}}. \label{eq:szcirc}
\end{eqnarray}
The SAM density points up (down) away from the phase singularity for RCP (LCP) and gradually becomes transverse as we approach it. At the singularity the electric field is linearly polarized along the $z-$direction, constituting a point of undefined spin orientation ($\bm{S}=0$). The SAM density lays on the transverse plane around this point and points azimuthally, completing a full $2\pi$ rotation around it. The resulting vector field covers the entire north (south) hemisphere of the SAM unit sphere for RCP (LCP) excluding the equator (theoretically covered at $r=0$ where the SAM vanishes). The textures are shown in Fig.~\ref{fig:F1} at a circle of radius $0.75\lambda$ around the phase singularity for $\ell=-1$ and $\sgz=1$ (a) and for $\ell=+1$ and $\sgz=-1$ (b). Hue color and brightness represent the azimuth and elevation in the SAM unit sphere respectively (see inset). Although the texture spans an entire hemisphere of the SAM unit sphere, it cannot be classified as a meron since at the center the vector field is not defined, i.e., it is a meronic spin defect with fractional topology~\cite{Wang2023PhotonicLoops,Wang2022TopologicalLight,Wu2025PhotonicMonopoles}. 
From Gauss's law, this localized texture of non-transverse polarization decorates the phase singularity of an ordinary optical vortex, similar to how vortices of circularly polarized waves turn into $C-$points when their phase singularity is decorated by an inhomogeneous transverse polarization~\cite{Angelsky2023PolarizationAspects}.
\begin{figure*}[t!]
    \centering
    \includegraphics[width=\linewidth]{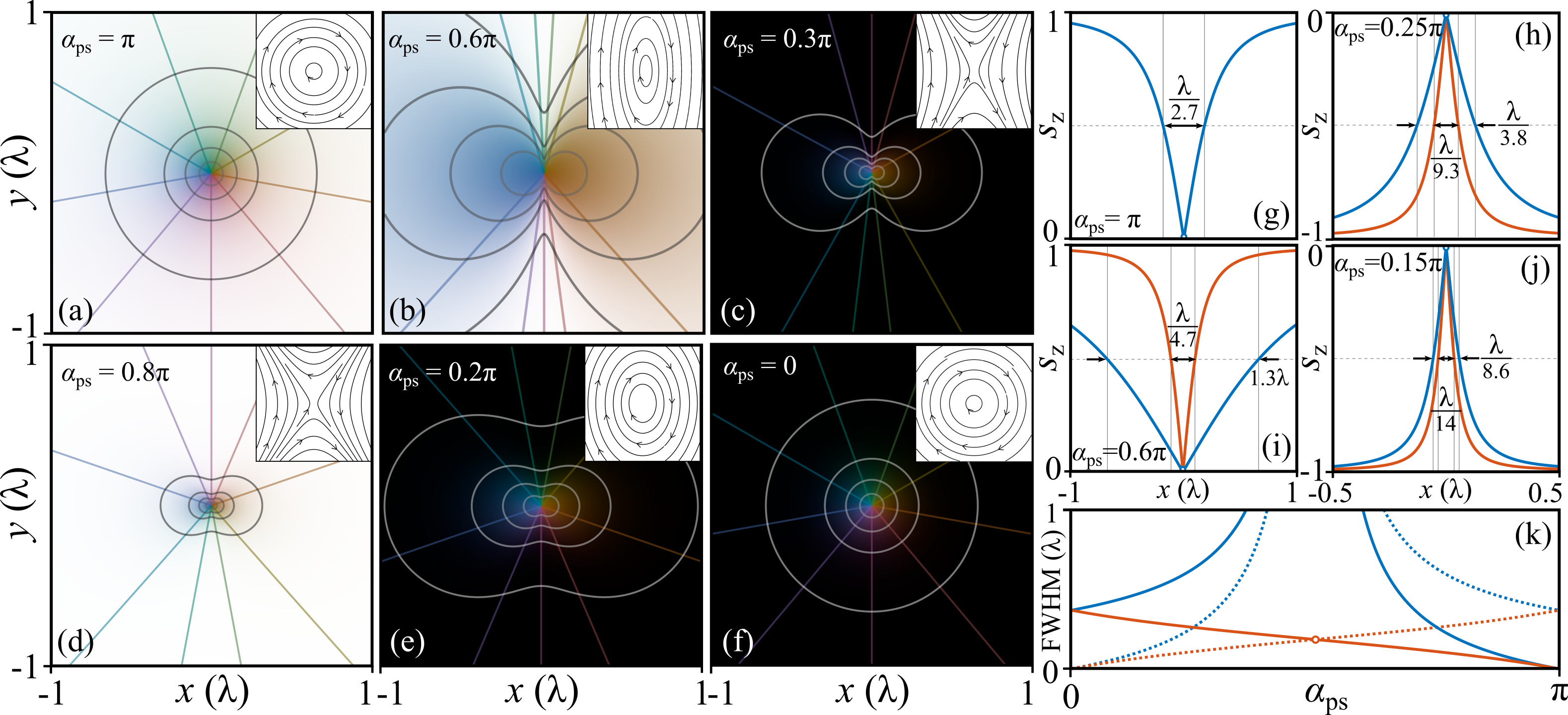}
    \caption{\textbf{Geometry of meronic spin defects around vortices.} SAM textures around the phase singularity of a vortex beam shown with hue-brightness color map as in (\ref{fig:F1}) with the contours of $\alpha_{\rm spin}$  (gray curves) and $\varphi_{\rm spin}$ (colorful lines), and the streamlines of the transverse SAM (inset). (a)-(b)-(c) correspond to $\ell=-1$, (d)-(e)-(f) to $\ell=+1$. (g)-(h)-(i)-(j) show the slices of $s_z$ along major (blue) and minor (red) axes of the transverse polarization ellipse for $\ell=-1$, including the FWHM. The value of $\apps$ for all previous plots is also shown. The FWHM along major (blue) and minor (red) axes as a function of $\apps$ is shown in (k) for $\ell=1$ (solid curves) and $\ell=-1$ (dotted curves).}
    \label{fig:F2}
\end{figure*}

\subsection{Anisotropic textures}
When the transverse polarization state is not circular $\lr{\sgz^2\neq1}$, meronic spin defects become highly anisotropic by virtue of the term $\tan\lr{\chi}$ in (\ref{eq:phsp}) and the factors in $\sgz^2$ in (\ref{eq:sz}). The contours of the spin azimuth (\ref{eq:phsp}) squeeze around the major axis of the transverse polarization ellipse (at $\phi=-\phps/2$), and the contours of the polar angle (\ref{eq:sz}) lose their circular symmetry. The topological features of the spin defect are invariant upon these deformations; one of them is the topological charge of the transverse SAM vortex (\ref{eq:phsp}) calculated as the circulation of (\ref{eq:phsp}) in a path surrounding the phase singularity. It follows immediately that the singularities of $\phsp$ are of the same order as the singularities of $\gamma$, and the ellipticity $\chi$ in (\ref{eq:sz}) only changes their relative sign. Since $\phps$ is constant and $\rrm{sgn}\lr{\chi}=\rrm{sgn}\lr{\sgz}$ we can express the circulation as 
\begin{equation}\label{eq:TCphsp}
    \frac{1}{2\pi}\oint\nabla\phsp\cdot\rrm{d}\bm{l}=\frac{\rrm{sgn}\lr{\sgz}}{2\pi}\oint\nabla\gamma\cdot\rrm{d}\bm{l}=-\rrm{sgn}\lr{\ell\sgz},
\end{equation}
where we have used the phase of the field ratio in (\ref{eq:ratio}). Eq. (\ref{eq:TCphsp}) determines the topology of the transverse SAM vortex, yielding circulations for $\rrm{sgn}\lr{\ell\sgz}=-1$ and saddles for $\rrm{sgn}\lr{\ell\sgz}=1$. Since a circulation cannot be continuously deformed into a saddle~\cite{Freund1995SaddlesFields}, there is a degenerate case (neither vortex nor saddle) which delimits them. This happens if the transverse polarization is linear $\sgz=0$, where the SAM density is entirely transverse and points orthogonally to the major axis of the transverse ellipse. The streamlines of the SAM density in a frame $x'y'$ aligned with the transverse polarization ellipse's major axis (a rotation of the $xy$ frame by $-\phps/2$) satisfy $\rrm{d}y'/\rrm{d}x'=\tan\lr{\phsp+\phps/2}$, and yield the family of curves 
\begin{equation}\label{eq:streamlines}
    -x'^2\frac{\rrm{sgn}\lr{\ell}}{\tan\lr{\chi}}+y'^2=C,
\end{equation}
with $C$ a constant. These are hyperbolas (saddles) for $\rrm{sgn}\lr{\ell\sgz}>0$ and ellipses (circulations) for $\rrm{sgn}\lr{\ell\sgz}<0$. The minor axis of the transverse polarization ellipse coincides with the major axis of an elliptical streamline, with its eccentricity given by $\sqrt{1-\tan^2\lr{\chi}}$. The streamlines degenerate into straight lines for linear polarization ($\chi=0$ and $\sgz=0$), where the topological charge of the SAM vortex is degenerate.

In addition to the anisotropy of the transverse SAM, the geometry of the out-of-plane component $s_z=\cos\lr{\apsp}$ changes with $\tan\lr{\chi}$, whose contours become a function of both radial and azimuthal coordinates. From (\ref{eq:ratio},~\ref{eq:sz}) the contours in the rotated frame read 
\begin{equation}\label{eq:contour_sz}
    k^2r^2=2\frac{\lr{1-\rrm{sgn}\lr{\ell}\sgz}\lr{1+\cos\lr{2\phi'}\sqrt{1-\sgz^2}}}{\sgz^2\lr{1/s_z^2-1}},
\end{equation}
with $\phi'=\phi+\phps/2$. Regardless of the value of $s_z$, we can evaluate the anisotropy of the $s_z$ contour as the ratio between the maximum $r_{\rm max}$ and minimum $r_{\rm min}$ radii in (\ref{eq:contour_sz}), obtained by maximizing and minimizing the angular term $\cos\lr{2\phi'}$ yielding
\begin{equation}\label{eq:squeezing_sz}
    \frac{r^2_{\rm max}}{r^2_{\rm min}}=\frac{1+\sqrt{1-\sgz^2}}{1-\sqrt{1-\sgz^2}}=\frac{1+\sin{\lr{\apps}}}{1-\sin{\lr{\apps}}},
\end{equation}
where the dependence of $s_z$ factors out. As expected Eq. (\ref{eq:squeezing_sz}) yields $r^2_{\rm max}=r^2_{\rm min}$ for circular polarization $\lr{\sgz^2=1}$, where the contours are circularly symmetric. We also obtain the full-width at half maximum (FWHM) of the $s_z$ distribution from Eq. (\ref{eq:contour_sz}) which reads
\begin{equation}
    % \rrm{FWHM}=\frac{\lambda}{\pi}\sqrt{\frac{2}{3}}\frac{\lr{1+\rrm{sgn}\lr{\ell}\cos\lr{\apps}}\lr{1\pm\sin{\lr{\apps}}}}{\cos^2\lr{\apps}},
    \rrm{FWHM}=\frac{\lambda}{\pi|\sgz|}\sqrt{\frac{2}{3}}\sqrt{1-\rrm{sgn}\lr{\ell}\sgz}\sqrt{1\pm\sin\apps},
\end{equation}
where upper and lower signs refer to the slices along major and minor axes of the transverse polarization ellipse. The transverse confinement of such structures only depends on the relation between spin and topological charges, $\sigma_z$ and $\ell$.

The anisotropic meronic spin defects are depicted in Fig.~\ref{fig:F2}: (a-f) show the SAM texture using hue-brightness color map as in Fig.~\ref{fig:F1}, with the streamlines of the transverse SAM shown at the top right corner. The corresponding transverse polarization state is given by $\apps$ as an inset (we set $\phps=0$ for simplicity), while top and bottom rows refer to $\ell=-1$ and $\ell=+1$ respectively. 
% since for $\phps\neq0$ the SAM structure simply undergoes a global rotation. 
The circularly symmetric texture for transverse circular polarization is shown in (a) for $\apps=\pi$ and $\ell=-1$, and (f) for $\apps=0$ and $\ell=+1$. The singularity $\bm{S}=0$ is surrounded by a circulating transverse SAM texture which gradually becomes longitudinal within a distance of the order of the wavelength $\lambda$. As the ellipticity of the transverse polarization increases ($\apps$ approaching $\pi/2)$, the textures are deformed but the topological charge of the SAM vortex is preserved, it becomes degenerate for $\apps=\pi/2$ and then swaps sign, with the streamlines changing their geometry from circulation to saddles and vice-versa. Equivalent behavior is obtained starting from $\ell=-1$ and $\apps=\pi$ and gradually decreasing $\apps$ towards $\apps=0$. By virtue of the spin-orbit factor in (\ref{eq:ratio}), when $\rrm{sgn}\lr{\ell}\sigma_z\to+1$ the meronic spin defect tends to a circularly symmetric texture with infinitesimally small size for $\sgz\to1$ and $\ell=1$, and for $\sgz\to-1$ and $\ell=-1$, as shown in (c) and (d). 

The squeezing of the out-of-plane component $s_z$ along the minor (red) and major (blue) axes of the transverse polarization ellipse is shown in panels (g-j) for different values of $\apps$, highlighting the FWHM normalized to the wavelength $\lambda$. The SAM density is highly transverse around the phase singularity $\lr{s_z\to0}$ and becomes longitudinal gradually away from it (g-h-j). For circular polarization with $\rrm{sgn}\lr{\ell}\sigma_z=-1$ (g) the two slices are equal, with the $s_z$ component tilting in a subwavelength fashion $\lr{\rrm{FWHM}=\lambda/2.7}$. This constitutes the maximum size for this kind of cylindrically symmetric spin textures, which is independent of the vortex beam shape and arises from Gauss's law. When the transverse polarization becomes more elliptical approaching linear states, the FWHM along the major axis increases, decreasing along the minor axis (i). At $\apps=0.5$ the FWHM of $s_z$ is not defined as the SAM is purely transverse (sagittal polarization). Then, as the transverse polarization approaches circular states with aligned spin and topological charges $\sigma_z\ell>0$, the texture becomes increasingly symmetric and both FWHM shrink (h), in the limit $\rrm{sgn}\lr{\ell}\sigma_z\to+1$ yielding two identical slices with infinitely narrow width $\rrm{FWHM}\to0$. The FWHM along the minor (red) and major (blue) axes is also plotted in (k) as a function of $\apps$ for $\ell=1$ (solid line) and $\ell=-1$ (dotted line). 

Here we have shown the anisotropy of meronic spin defects around vortices induced from the ellipticity of the transverse polarization, and not from the anisotropy of the vortex phase itself. In a more general case, the isotropy of the SAM defect will also be altered if the vortex beam has an elliptical anisotropy ellipse~\cite{Dennis2004LocalTwirl}, yet the geometry of the contours will differ, and the field ratio (\ref{eq:ratio}) will lose its cylindrical symmetry as shown in the Appendix (\ref{sec:A5}). 

\subsection{Topological charge}\label{sec:TC}
The SAM defect covers a fractional area of the SAM unit sphere, whose solid angle is calculated as a contour integral in the transverse plane~\cite{McWilliam2023TopologicalMulti-skyrmions}
\begin{equation}\label{eq:Q}
    Q=\frac{1}{4\pi}\oint_\Gamma s_z\nabla\phsp\bm{\rrm{d}l}=\frac{1}{2}\lr{\sum_j s_z^{(j)}q^{(j)}-s_z^{(\infty)}q^{(\infty)}}.
\end{equation}
The closed path $\Gamma$ avoids all $j$ singularities of the spin azimuth with topological charge $q^{(j)}=(2\pi)^{-1}\oint\nabla\phsp\rrm{d}\bm{l}=(2\pi)^{-1}\rrm{sgn}\lr{\sgz}\oint\nabla\gamma\rrm{d}\bm{l}$, while $s_z^{\lr{\infty}}$ and $q^{(\infty)}$ denote the $z$-component spin at the boundary of the texture $\lr{r/\lambda\gg1}$ and its topological charge~\cite{McWilliam2023TopologicalMulti-skyrmions}. The singularities appear at the vortices of the transverse and longitudinal fields, where the azimuth $\phsp$ becomes undefined (\ref{eq:phsp}). Splitting (\ref{eq:Q}) in terms of transverse $\lr{\perp}$ and longitudinal $\lr{\parallel}$ field vortices weighted by their corresponding spin value $s_z$ around them we obtain the expression 
\begin{equation}\label{eq:Qsimplified}
    Q=
    % \frac{1}{2}\lr{\sum_{||} s_z^{||}q^{||}+\sum_{\perp} s_z^{\perp}q^{\perp}-s_z^{(\infty)}q^{(\infty)}}=
    \frac{1}{2}\lr{\sum_{||} s_z^{||}q^{||}-s_z^{(\infty)}q^{(\infty)}},
\end{equation}
where the summation runs over all zeros of $\psi_z$, and the zeros of $\psi_\perp$ have not been included in (\ref{eq:Qsimplified}) since from (\ref{eq:sz}) $s_z=0$. We can simplify $s_z^{(\infty)}=\rrm{sgn}\lr{\sgz}$, since for a paraxial field the SAM density becomes longitudinal far away from the origin, where the transverse field dominates [as seen from Eq. (\ref{eq:ratio}) for  $r/\lambda\gg1$ and Eq. (\ref{eq:sz}) for $\rho\to0$]. This guarantees the solid angle $Q$ to be quantized, as we show below.

\begin{figure*}[t!]
    \centering
    \includegraphics[width=0.8\linewidth]{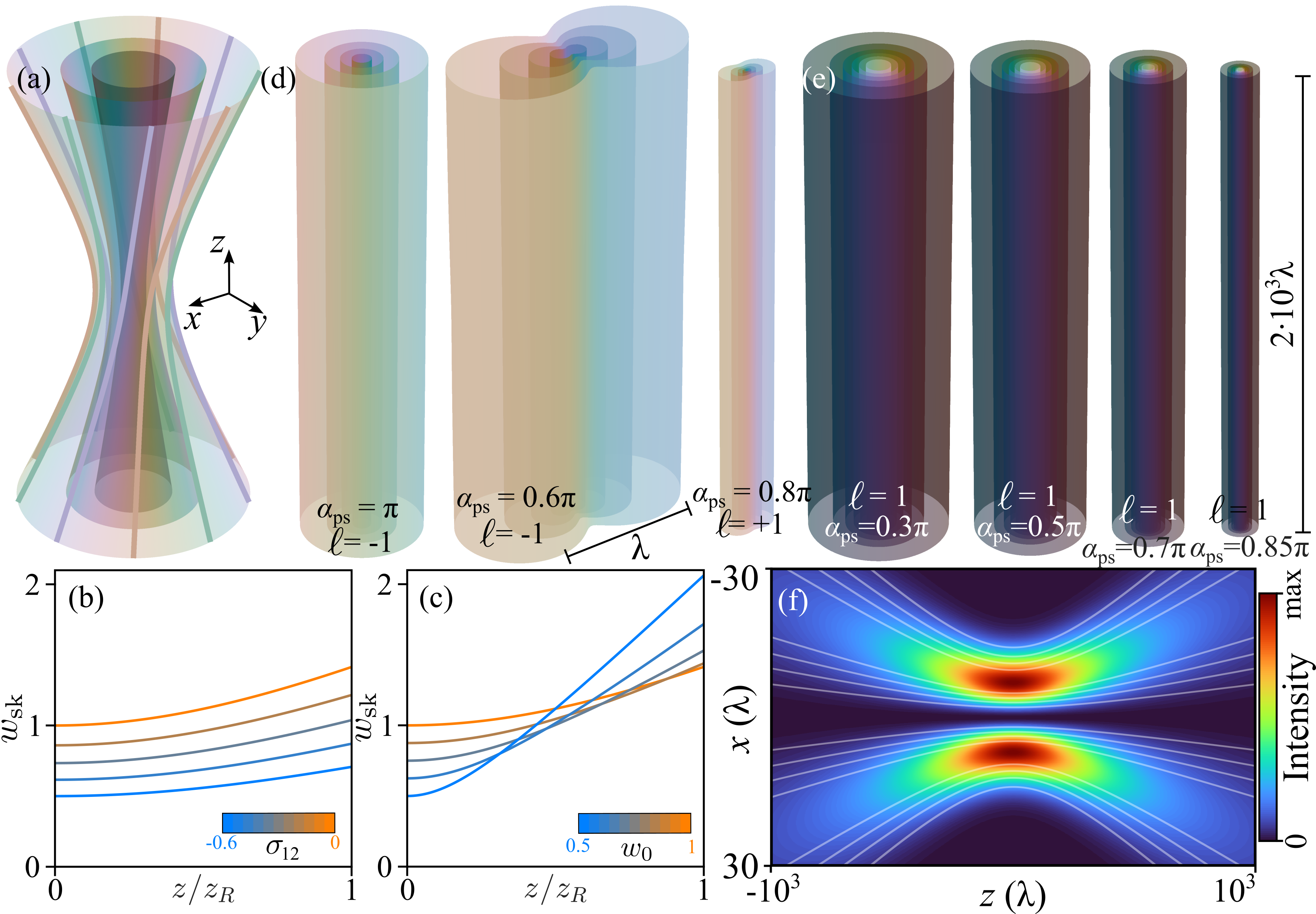}
    \caption{\textbf{Propagation of polarization textures.} (a) Polarization contours in Stokes skyrmions of transverse polarization, (b) the evolution of their characteristic size $w_{sk}$ as a function of the propagation distance $z/z_R$ for fixed beam width $w_0$ and decreasing $\sigma_{12}$, and (c) for fixed $\sigma_{12}$ and decreasing beam waist $w_0$. (d) Non-spreading SAM contours in a vortex beam with uniform transverse polarization for different values of $\apps$. (e) Same as (d) for the TA-PS polarization contours. The total propagation distance in (d)-(e) is $2\cdot10^3\lambda$, and the evolution of the vortex beam is shown in (f), the white curves show the contours $r\propto w(z)$. The parameters of the vortex beam (\ref{sec:A3}) are $w_0=10\lambda$, $\ell=1$, $p=0$. The hue-brightness color maps as in Fig.~\ref{fig:F1} show the spherical coordinates in the corresponding parametric sphere: (a) PS, (d) SAM unit sphere, (e) TA-PS.}
    \label{fig:F3}
\end{figure*}
The topological charges of the SAM vortices are organized by the TA field ratio (\ref{eq:ratio}) yielding $\gamma=\pi/2-\rrm{sgn}\lr{\ell}\phi$. At the boundary reads $q^{(\infty)}=+\rrm{sgn}\lr{\sgz}\oint\nabla\gamma\rrm{d}\bm{l}/2\pi=-\rrm{sgn}\lr{\sgz\ell}$. Finally, the vortices of the longitudinal field contribute to the solid angle (\ref{eq:Qsimplified}) as $q^{||}=+\rrm{sgn}\lr{\sgz}\cdot\rrm{sgn}\lr{\ell}\cdot\lr{|\ell|-1}$, where we have used the phase structure of $\psi_z$ in (\ref{eq:psi_z}) and each of the vortices is weighted by $s^{||}_z=\rrm{sgn}\lr{\sgz}$. Plugging the charges in (\ref{eq:Qsimplified}) we obtain the fractional solid angle of the meronic spin defect as
\begin{equation}\label{eq:Qfrac}
    Q_{\rm spin}=\frac{\ell}{2}\rrm{sgn}^2\lr{\sgz},
\end{equation}
where we have kept the function $\rrm{sgn}^2\lr{\sgz}$ yielding $+1$ except for $\sgz=0$ where the texture is degenerate and the solid angle covered in the unit sphere is zero. Although both $\sgz\gtrless0$ have the same solid angle for a given $\ell$, the spin textures for $\sgz\ell<0$ are circulations, while for $\sgz\ell>0$ they are saddles, as shown in Fig.~\ref{fig:F2}. 

In the derivation we have assumed that transverse field zeros do not contribute to (\ref{eq:Qsimplified}) which is valid as long as transverse and longitudinal field vortices do not coalesce. This situation is stable upon perturbation of the transverse scalar field $\psi_\perp$. The solid angle of the unstable SAM defect is bounded $Q_{\rm spin}^\star=\rrm{sgn}\lr{\ell}/2$, similarly as for the TA-Stokes texture $Q_{z\perp}^\star=\rrm{sgn}\lr{\ell}$ (\ref{eq:Qsk_zt}). Repeating the calculation for the TA-PS Stokes texture with unfolded phase singularities we obtain a stable skyrmion number $Q_{z\perp}=\ell$, in contrast to the result reported in~\cite{Mata-Cervera2025SkyrmionicVortices}. Both $Q_{\rm spin}^{\star}$ and $Q_{z\perp}^{\star}$ are not preserved upon arbitrarily small perturbation, while $Q_{\rm spin}$ and $Q_{z\perp}$ are more general and preserved upon perturbation of the vortex core~\cite{Mata-Cervera2026UnfoldingTextures}.

\section{Propagation}
\subsection{Ideal propagation and confinement}
These topological features have been discussed for a certain propagation plane $z=z_0$, where the vortex beam has a characteristic size $w(z_0)$. Surprisingly, as it propagates away from this plane, the normalized polarization and spin structures remain confined even when transverse (\ref{eq:psi_perp}) and longitudinal (\ref{eq:psi_z}) fields both diffract, but the resulting polarization field (\ref{eq:ratio}) is itself non-spreading. This contrasts from previously known non-diffractive wave modes, since here the polarization texture does not spread while the electromagnetic field spreads. Such novel wave phenomenon was first revealed in a seminal paper by Afanasev et al.~\cite{Afanasev2023NondiffractiveBeams}, which considered the non-spreading nature of the modulus of (\ref{eq:ratio}) and its relation to the 3D polarization matrix~\cite{Varshalovich1988QuantumMomentum}, later similar behavior was found at far-field zeros of electromagnetic radiation~\cite{Vernon2024Non-diffractingRadiation} and at acoustic vortex lines~\cite{Kille2024NondivergentNodes}. Recent works reported the experimental observation of the associated non-spreading skyrmionic textures at optical~\cite{Mata-Cervera2026Diffraction-FreeVortices} and acoustic vortices~\cite{Annenkova2025UniversalSound}. 

As a polarization texture, it contrasts with previously known structures of spatially-varying polarization such as paraxial skyrmionic beams~\cite{Gao2020ParaxialBeams,Beckley2010FullBeams}. In the latter, lines of constant polarization are contained in hyperboloids in real space, see Fig.~\ref{fig:F3}(a). Thus the polarization texture expands as the beam propagates, and the lines of constant polarization flow away from the propagation axis. This can be explained from the ubiquitous rule governing the behavior of vortices with topological charge $\ell$, $\psi_\ell\propto\lr{r/w}^{|\ell|}e^{i\ell\phi}$. If we construct a polarization texture from two vortices with $\ell'\neq\ell$ and different polarizations, at any point in the transverse plane $r_0$ the ratio between their amplitudes will become propagation-dependent $|\psi_\ell/\psi_{\ell'}|\lr{r_0}\propto1/w^{|\ell|-|\ell'|}$ since the vortices themselves diffract $w=w\lr{z}$.

Yet, the longitudinal field naturally associated with a single vortex beam (\ref{eq:psi_z}) breaks this fundamental conception $\psi_z\propto\lr{r/w}^{|\ell|}e^{i\lr{\ell-1}\phi}/r$, and surprisingly its amplitude near the vortex core decays in propagation at the same rate as the transverse field (\ref{eq:ratio}), even though their topological charges are different. As a result, the TA polarization textures around a vortex line trace out non-spreading polarization contours in propagation. The geometry of the TA polarization contours differs depending on the space chosen to represent it, i.e., the meridians in the SAM unit sphere (constant $s_z$) trace out cylinders with asymmetric cross-section~\ref{fig:F3}(d) while the meridians of the TA-PS (constant $\rho$) are circular cylinders (e), but in any case these surfaces do not expand. 

As other kinds of polarization textures, their size can be controlled by changing the intensity balance of the constituent vortex components. Take for example a paraxial skyrmionic beam, whose size (\ref{sec:A4}) is $w_{\rm sk}=1$ at $z=0$ when both RCP and LCP waves have identical weight. The skyrmion size $w_{\rm sk}$ increases with propagation, and its divergence is determined by the beam waist $w_0$. Nonetheless, $w_{\rm sk}$ can be shrunk down to arbitrary small (subwavelength) scales by decreasing the weight of the vortex of lower order. Shrinking the size in this way does not modify the divergence of the skyrmionic texture~\ref{fig:F3}(b), in contrast to doing so by decreasing the beam waist $w_0$, which ultimately increases the divergence of the polarization texture (c). Similarly, the TA polarization texture around vortices can be shrunk by reducing the peak intensity of the scalar component of lowest order, in this case, by attenuating the longitudinal component. By virtue of the square root term in (\ref{eq:psi_z}), the longitudinal field $\psi_z$ is weakened with respect to the transverse field $\psi_\perp$ making the transverse polarization circular with the spin charge parallel to the topological charge, $\rrm{sgn}\lr{\ell}\sgz\to+1$, as shown in (d) and (e). Both the SAM and TA-Stokes contours tend to form infinitesimally thin skyrmionic tubes with constant circular cross-section in the limit $\rrm{sgn}\lr{\ell}\sigma_z\to+1$. The evolution of the vortex intensity profile in the $xz$ plane is shown in (f). Note that even though both transverse (\ref{eq:psi_perp}) and longitudinal (\ref{eq:psi_z}) electric field envelopes are paraxial, thus they vary at scales much larger than the wavelength, their combined polarization ellipse changes orientation at scales well below the wavelength. As it happens with linear momentum~\cite{Dennis2008SuperoscillationPatterns}, spin~\cite{Du2019Deep-subwavelengthMomentum} or transverse polarization~\cite{Maxwell2025StochasticFields}, quantities derived from paraxial wave fields may locally vary much faster than the fields themselves.

\subsection{Perturbed propagation}
In real scenarios, an unstable $\ell$-th order vortex beam unfolds into $|\ell|$ stable phase singularities of order $\rrm{sgn}\lr{\ell}$. The Poincar\'e index is preserved in the splitting, as the high-vortex of order with index $+1$ is unfolded into $|\ell|$ vortices with index $+1$, accompanied by $\lr{|\ell|-1}$ saddles with index $-1$~\cite{Freund1995SaddlesFields,Dennis2009SingularSingularities,Dennis2006RowsBeam}. For large-scale polarization textures this effect is usually negligible, as the separation between the vortices barely affects the overall polarization structure, and the emerging polarization singularities are too closely separated to be observed experimentally. However, the polarization textures discussed here occupy remarkably small areas of the order of the wavelength $\lambda$ or smaller, so that the singularity splitting plays an important role. 
\begin{figure}[t!]
    \centering
    \includegraphics[width=\linewidth]{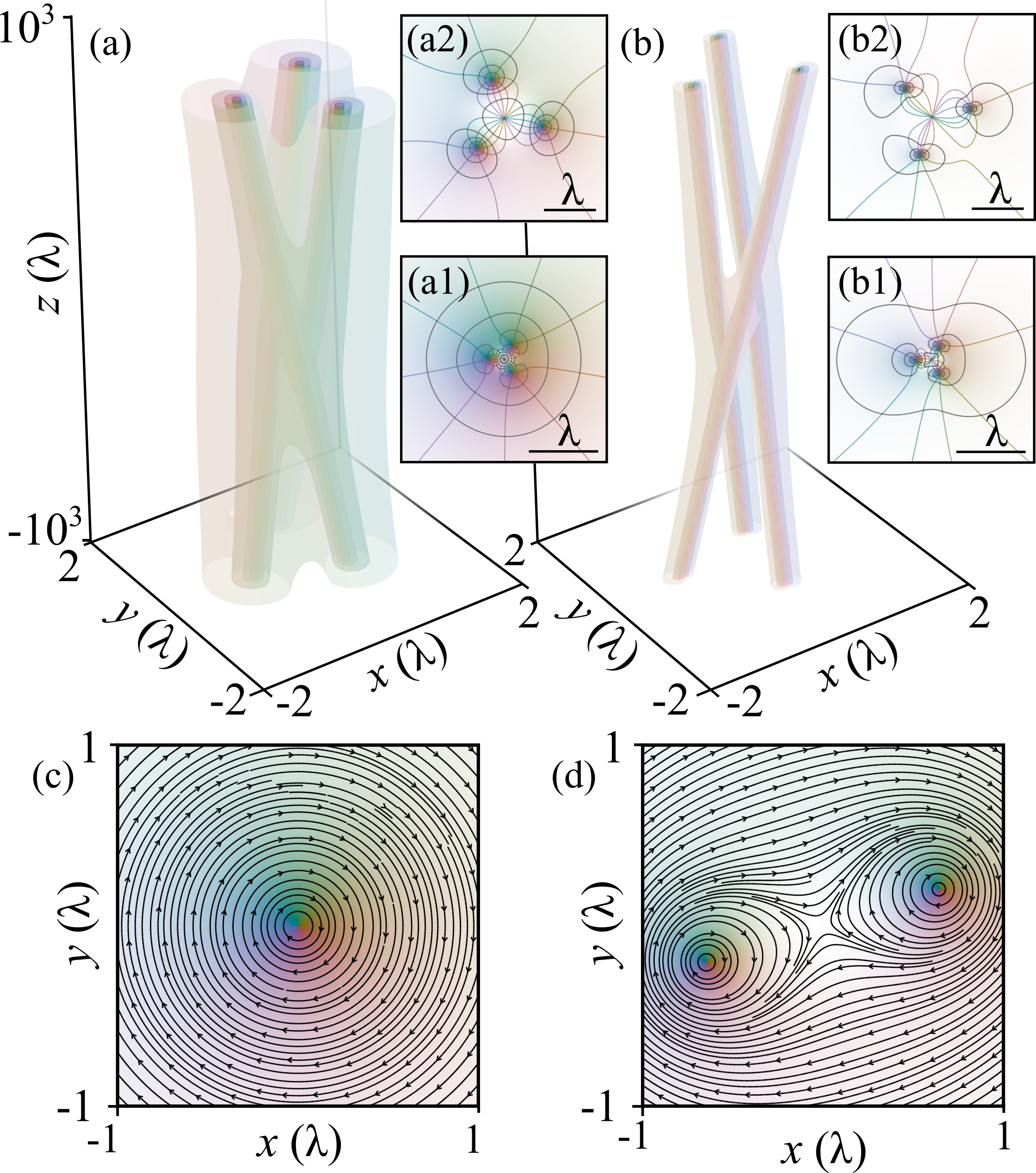}
    \caption{\textbf{Propagation of unfolded high-order spin meronic defects around vortices.} (a) shows the spin meronic defect in a vortex beam with $\ell=-3$ and $\sgz=+1$ ($\apps=\pi$), the insets (a1) and (a2) show the spin texture at the planes $z=0$ and $z=10^3\lambda$, and the same for $\ell=3$ and $\apps=0.8\pi$ shown in (b) with the corresponding textures (b1) and (b2). The parameters of the vortex beam (\ref{sec:A3}) are $w_0=10\lambda$, $\ell=3$, $p=0$. Ideal spin meronic defect around a vortex with $\ell=-2$ and $\sgz=+1$ (c) and the same defect upon perturbation of the vortex (d), the black arrows are the streamlines of the transverse SAM.}
    \label{fig:F4}
\end{figure}

Fig.~\ref{fig:F4}(c) shows a ideal spin defect in a vortex beam of charge $\ell=-2$ and $\sgz=1$, which occupies a significantly larger area than the fundamental $\ell=1$ case [Fig.~\ref{fig:F2}(a)] by virtue of a slower decay of $s_z$ (\ref{eq:szcirc}). The streamlines of the transverse spin are depicted by black arrow curves. When the vortex field is slightly perturbed, the second-order spin defect unfolds into two fundamental spin defects (d), each of them with similar size as in Fig.~\ref{fig:F2}(a). The streamlines of transverse spin unfold into two circulations separated by one saddle yielding a standard singularity unfolding pattern (d), so that the topological charge (\ref{eq:TCphsp}) is conserved. Yet, the solid angle in the spin unit sphere of the unfolded texture is twice the solid angle of the ideal texture as discussed in Sec.~\ref{sec:TC}.

Splitting of high-order vortices can be modeled by disturbing the vortex beam with a weak wave $\varepsilon$~\cite{Dennis2006RowsBeam,Berry2001KnottedWaves,Berry2001KnottingSpacetime}, modifying (\ref{eq:psi_perp}) into 
\begin{equation}\label{eq:perturbation}
    \psi_\perp=\lr{\frac{r}{w(z)}}^{|\ell|}e^{i\ell\phi}+\varepsilon.
\end{equation}
The perturbation which splits the zeros of the field into a circle of radius $r_n$ determined by the perturbation amplitude, and the azimuthal position $\phi_n$ given by its phase 
\begin{eqnarray}
    r_n=w(z)|\varepsilon|^{1/|\ell|},\label{eq:r_pert}\\
    \phi_n=\frac{\arg{\lr{\varepsilon}}+\pi}{\ell}+n\frac{2\pi}{\ell}.\label{eq:phsp_pert}
\end{eqnarray}
Expressing the transverse coordinate as a complex number $\zeta=re^{i\phi}$ the field becomes a function of complex input $\psi_\perp=\psi_\perp(\zeta)$ which vanishes at $\zeta=\zeta_n=r_ne^{i\phi_n}$ (\ref{eq:r_pert}, \ref{eq:phsp_pert}). Expanding the transverse field (\ref{eq:perturbation}) around its zeros $\zeta_n$ using the binomial theorem we obtain
\begin{equation}\label{eq:expansion_zero}
    \psi_\perp(\zeta_n+\delta)\approx\frac{|\ell|}{w^{|\ell|}(z)}|\delta|e^{i\rrm{sgn}\lr{\ell}\arg\lr{\delta}}+O(|\delta/\zeta_n|^2),
\end{equation}
where $\delta$ is a small displacement around a zero of (\ref{eq:perturbation}), with the radial displacement given by $|\delta|$ and the azimuthal rotation by $\arg\lr{\delta}$. To first order in $|\delta/\zeta_n|$, the field  behaves locally as a first-order vortex around each of the unfolded zeros, $\psi_\perp\propto r'e^{i\rrm{sgn}\lr{\ell}\phi'}$ where $r'$ and $\phi'$ are the local polar coordinates at the position of the zero. The approximation becomes more exact as the distance between the zeros is larger. Only then we can consider each of the singularities as ``independent" spin defects, while if the separation between them is smaller, they will show a joint spin texture. 

The SAM texture of the perturbed higher-order vortex beam will inherit the non-spreading features of a first-order vortex (\ref{eq:ratio}) locally around each of the unfolded phase singularities, yet each of the emerging spin meronic defects will propagate along a different trajectory. We show their propagation in Fig.~\ref{fig:F4}(a) for weakly-perturbed $\ell=-3$ vortex beam with LCP. Since $\rrm{sgn}\lr{\ell}\sgz=-1$, the size of the spin defect is maximum, and the three phase singularities yield a notable joint texture at the waist plane where the separation between singularities is minimum (a1). Then they spread away from each other as the vortex beam propagates (a2) but in any case keeping similar transversal size. Their joint spin texture becomes almost unnoticeable if each of the individual spin textures shrinks, for instance for $\ell=3$ and $\apps=0.8\pi$, thus the interaction between them is negligible (b1, b2). In any case as the vortex propagates to the far-field and the phase singularities spread away, each of them carry a meronic spin defect which is confined in a non-spreading cylinder. The asymmetry of the cylinder cross-section determined by the factor $\rrm{sgn}\lr{\ell}\sgz$. The vortex beam and diffraction distance used in Fig.~\ref{fig:F4}(a-b) are identical to that of Fig.~\ref{fig:F3}(d-f).
\subsection{Turbulence}
The intrinsic confinement of the texture of TA polarization not only manifests as a non-spreading propagation in free-space, but also provides the texture with stronger robustness under random perturbations compared to standard Stokes skyrmionic beams, whose confinement relies on the precise control over phase, amplitude and polarization of light. To demonstrate this we propagate light through a turbulent channel, implemented with a random phase mask, and we analyse the polarization distribution at the far-field. The phase mask is obtained using an inverse Fourier transform with the Von-K\'arm\'an power spectral density, details can be found in~\cite{Goodman2015StatisticalOptics,Schmidt2010NumericalMATLAB}. The coherence length $r_0$ determines the distance at which two points in the phase screen are uncorrelated, and its quotient $r_0/w_0$ with the beam width $w_0$ serves as a measure of the turbulence strength. 

We analyse two cases with weak $r_0/w_0=5.75$ and strong $r_0/w_0=1.5$ turbulence disturbance. For a legitimate analysis we compare two structures with the same ideal skyrmion number ($Q=1$): a Stokes skyrmion constructed from a Gaussian and a first-order LG vortex beam, and the TA skyrmionic texture around a circularly polarized first-order LG vortex beam. For each random realization we integrate the solid angle $Q=(4\pi)^{-1}\iint_\Omega \bm{s}\cdot\lr{\partial_x\bm{s}\times\partial_y\bm{s}}\rrm{d}x\rrm{d}y$ spanned by the normalized Stokes and TA-Stokes vectors $\bm{s}\lr{x,y}$ within a finite region $\Omega$ of the transverse plane. Given the intrinsic confinement of the TA texture, defining the integration region $\Omega$ as a disk of radius $3\lambda$ around the vortex core suffices. For a standard Stokes skyrmion, the severe deformations introduced by the turbulence mask require the identification of $\Omega$ to be done with a more sophisticated active contour method~\cite{Kass1988Snakes:Models}, details and numerical codes can be found in Sec. 5 of Ref.~\cite{Peters2026ExtractingTutorial}. 

\begin{figure}[t!]
    \centering
    \includegraphics[width=\linewidth]{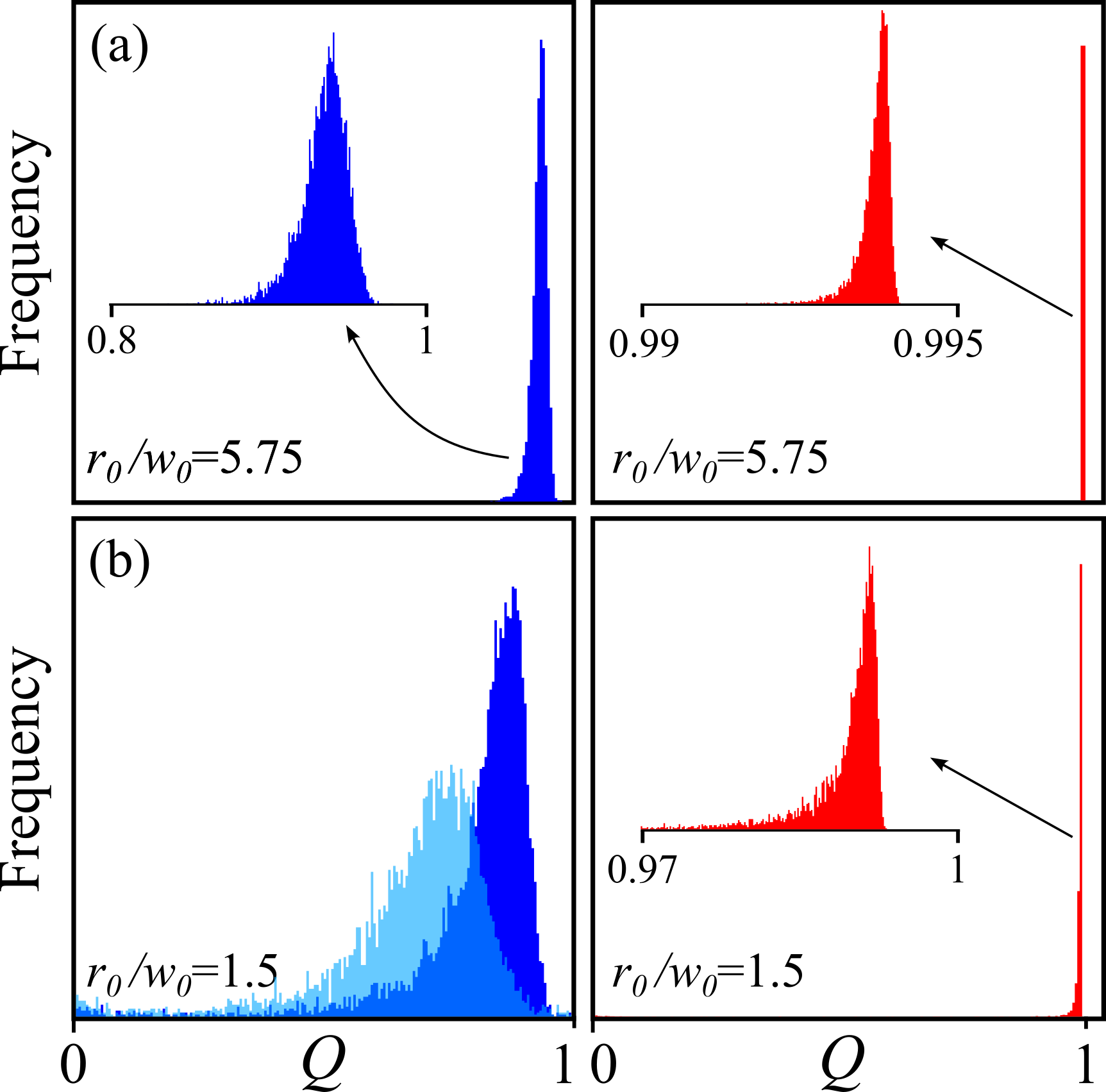}
    \caption{\textbf{Propagation through atmospheric turbulence.} Histograms of the solid angle $Q$ for a standard skyrmionic texture (left) and a skyrmionic texture of TA polarization (right) for two different turbulence strengths $r_0/w_0=5.75$ (a) and $r_0/w_0=1.5$ (b). Dark (light) blue refer to standard textures where the vortex beam $\ell=1$ carries 5 (2) times the power of the Gaussian beam $\ell=0$. The TA texture shows significantly stronger robustness upon perturbation than the standard skyrmionic texture. The results are obtained for 6000 random realizations of the turbulence mask.}
    \label{fig:F5}
\end{figure}

The histograms of the solid angle $Q$ for 6000 random realizations are shown in Fig.~\ref{fig:F5} for weak $r_0/w_0=5.75$ (a) and strong turbulence $r_0/w_0=1.5$ (b) comparing a standard Stokes skyrmion (left) and TA Stokes skyrmion (right). The results clearly show stronger robustness of the TA texture, which maintains a distribution of near-integer $Q$ (up to truncation error) even in stronger turbulence, in contrast with the standard Stokes texture whose $Q-$distribution broadens significantly and its maximum shifts towards smaller values of $Q$. Moreover, the performance of the latter depends strongly on the relative weight between the vortex $\ell=1$ and Gaussian $\ell=0$ beams (their powers $P_1/P_0$), since the truncated skyrmion number $Q$ is no longer a topological quantity and thus depends on how fast the Stokes vector converges towards its ideal value at infinity. As shown in (b)-left, having similar weights $P_1=2P_0$ (light blue) is even more fragile than having very disparate weights $P_1=5P_0$ (dark blue), yet the latter complicates its experimental characterization.

The improved robustness of the TA texture can be attributed to its intrinsicalness to the vortex core, where the skyrmionic texture forms automatically from Gauss's law, thus it does not rely on how two polarized waves are judiciously shaped. In a standard Stokes texture, convergence of $Q$ to integer values requires the higher-order vortex to dominate over the lower-order one away from the origin, a condition that is satisfied ideally by asymptotic dominance of one mode over the other while their amplitudes decay to zero. Nonetheless, this asymptotic dominance breaks down when the decay of their amplitude profiles is distorted by the external perturbation.

\section{Discussion}
We have shown the topological structure of non-transverse polarization and transverse spin that arises naturally at scalar vortices around their phase singularity. The polarization texture presents skyrmionic topology in a parametric space describing the relative phases and amplitudes between transverse and longitudinal fields akin to the Poincar\'e sphere and associated Stokes skyrmionic beams. On the other hand, the resulting distribution of transverse spin angular momentum constitutes a meronic defect with fractional topology. The spin meronic defect presents highly anisotropic features which are result of the ellipticity of the transverse polarization, but such anisotropy is not present in the parametric representation of the polarization texture. 

These topological structures are strongly confined below the wavelength of light, product of the weak amplitude of longitudinal fields in the paraxial regime. The same subwavelength confinement is maintained regardless of the size of the vortex beam, its convergence or divergence and the propagation distance. Similar behaviour is found when the vortex beam is perturbed and the phase singularities split, yet each of the split singularities will give rise to their own non-spreading polarization texture. As propagating counterparts, these textures complement previously reported spin structures such as spin defects~\cite{Wang2022TopologicalLight}, localized around a point in 3D space; three-dimensional spin quasi-particles~\cite{Wang2025PhotonicSkyrmion,Wu2025PhotonicMonopoles,Wang2023PhotonicLoops}, distributed in a 3D volume; and evanescent topological spin structures~\cite{Fang2025TopologicalFields,Du2019Deep-subwavelengthMomentum}, localized around material interfaces.

This kind of non-transverse polarization textures notably differ with respect to standard Stokes skyrmionic textures. Breaking with previously known properties of polarization textures, which are intrinsically unbounded and diffractive~\cite{Ye2024TheorySkyrmions}, the structures shown here present subwavelength confinement around vortex lines and remain non-spreading during free-space propagation. Moreover, their size is fixed by constituent relations---Gauss's law which couples different field polarization components---in contrast with Stokes skyrmionic beams where the confinement of the texture is not guaranteed by any natural law. Second, Gauss's law and the paraxiality of the wave guarantee the convergence to a transverse polarization state within few wavelengths away from the phase singularity. The same behaviour could be maintained for different kinds of perturbations applied onto the field profile unless strong focusing is performed. This contrasts drastically with standard Stokes skyrmionic textures where the polarization far away from the origin does not converge to a fixed state in the presence of perturbation. As a result of the intrinsic strong localization, these textures approximately feature integer wrappings of their parametric space, the wrapping number changing by quantized amounts when new singularities are introduced in the field. As we have shown, such features provides these textures with significantly stronger robustness than standard Stokes polarization textures when propagating through a random turbulent channel. Yet, as linear fields propagating in free-space, these textures are not topologically protected in the strict sense~\cite{Maxwell2025StochasticFields}.

It remains unexplored how these structures extend to the non-paraxial regime, where the transverse polarization becomes inhomogeneous and electric and magnetic spins differ. Spin merons with non-spreading propagation around vortex lines have recently been demonstrated in acoustic waves both theoretically and experimentally~\cite{Annenkova2025UniversalSound}. These topological structures could in principle find counterparts in other systems such as surface~\cite{Smirnova2024Water-WaveSkyrmions} or quantum waves~\cite{Matthews1999VorticesCondensate}, where subwavelength confinement without spreading might find different applications. In optical systems, super-resolution techniques based on vortex phase singularities have achieved picometric metrology and object localization well below the diffraction limit~\cite{Yuan2019DetectingMetrology,Huang2009Super-ResolutionWaves}. It is a direction to explore whether the precision of these metrology techniques could be controlled by the non-transverse polarization texture arising around the singularity, where the spin structure can be confined to arbitrarily small scales.

\textbf{Data availability.} The data that support the findings of this manuscript are openly available at \href{https://doi.org/10.21979/N9/QHR47O}{this link}.

\textbf{Acknowledgements.} Y.S. Acknowledges support from Nanyang Technological University Start Up Grant, Singapore Ministry of Education (MOE) AcRF Tier 1 grants (RG147/23 and RT11/23), Singapore Agency for Science, Technology and Research (A*STAR) (M24N7c0080 and H25-MRO3489). M.A.P. acknowledges support from the Spanish Ministry of Science and Innovation, Gobierno de España, under Contract No. PID2021-122711NB-C21.

\textbf{Conflict of interest.} The authors declare no competing interests.
\section{Appendix}
\subsection{Spin angular momentum density}\label{sec:A1}
The electric field in cartesian basis is
\begin{equation}
    \bm{\psi}=\frac{\psi_\perp}{\sqrt{2}}
\begin{pmatrix}
    \lr{\cos\lr{\frac{\apps}{2}}e^{-i\frac{\phps}{2}}+\sin\lr{\frac{\apps}{2}}e^{i\frac{\phps}{2}}}\\
    -i\lr{\cos\lr{\frac{\apps}{2}}e^{-i\frac{\phps}{2}}-\sin\lr{\frac{\apps}{2}}e^{i\frac{\phps}{2}}}
    \\
    \sqrt{2}\rho e^{i\gamma}
\end{pmatrix},
\end{equation}
from which we obtain the SAM density $\bm{S}=\rrm{Im}\lr{\bm{\psi}^{*}\times\bm{\psi}}$ as
\begin{widetext}
\begin{equation}\label{eq:S_xyz}
\bm{S}=|\psi_\perp|^2
\begin{pmatrix}
    2\rho\left[\cos\lr{\apps/2}\lr{\cos\lr{\phps/2}\cos\gamma-\sin\lr{\phps/2}\sin\gamma}-\sin\lr{\apps/2}\lr{\cos\lr{\phps/2}\cos\gamma+\sin\lr{\phps/2}\sin\gamma}\right]\\
    2\rho\left[\sin\lr{\apps/2}\lr{\sin\lr{\phps/2}\cos\gamma-\cos\lr{\phps/2}\sin\gamma}-\cos\lr{\apps/2}\lr{\sin\lr{\phps/2}\cos\gamma+\cos\lr{\phps/2}\sin\gamma}\right]\\
    -\cos\lr{\apps}
\end{pmatrix}.
\end{equation}
\end{widetext}

\subsection{Spin direction for circular polarization}\label{sec:A2}
If the transverse polarization is circular, its major axis $\phps$ becomes undefined, but from the structure of TA-polarization (\ref{eq:ratio}) the spin azimuth reads
$$\phsp=-\frac{\phps}{2}    \blue{-}\rrm{sgn}\lr{\ell\cdot \sgz}\lr{\frac{\phps}{2}+\phi}-\frac{\pi}{2}.$$
Since we only consider the case $\rrm{sgn}\lr{\ell\cdot \sgz}=-1$ (otherwise the polarization texture is trivial and transverse) then the SAM azimuth becomes
$$\phsp=\phi-\frac{\pi}{2},$$
so the transverse SAM density points azimuthally.

\subsection{Laguerre-Gaussian beams}\label{sec:A3}
Through this work, we have used the Laguerre-Gaussian (LG) modes, solution to the paraxial wave equation in cylindrical coordinates $r=\sqrt{x^2+y^2}$, $\varphi=\tan^{-1}\lr{y/x}$ given by
$$\rrm{LG}_p^\ell(r,\phi,z)=\sqrt{\frac{2p!}{\pi(|\ell|+p)!}}\frac{1}{w(z)}\lr{\frac{\sqrt{2}r}{w(z)}}^{|\ell|}e^{-r^2/w^2(z)}\times$$
\begin{equation}\label{eq:LGbeam}
    L_p^{|\ell|}\lr{\frac{2r^2}{w^2(z)}}e^{i\lr{kr^2/2R(z)+\ell\phi+kz+\mathcal{G}(z)}}
\end{equation}
where $\ell$ and $p$ denote azimuthal and radial indices respectively, $w(z)=w_0\sqrt{1+\lr{z/z_R}^2}$ is the beam width with $w_0$ the beam waist, $z_R=kw_0^2/2$ the Rayleigh range and $k=2\pi/\lambda$ the wavenumber for a given wavelength $\lambda$. $\mathcal{G}(z)=-(2p+|\ell|+1)\tan^{-1}{\lr{z/z_R}}$ is the Gouy phase and $R(z)=z[1+\lr{z_R/z}^2]$ is the wavefront radius of curvature.

\subsection{Width of Stokes skyrmions}\label{sec:A4}
\begin{figure}[b!]
    \centering
    \includegraphics[width=\linewidth]{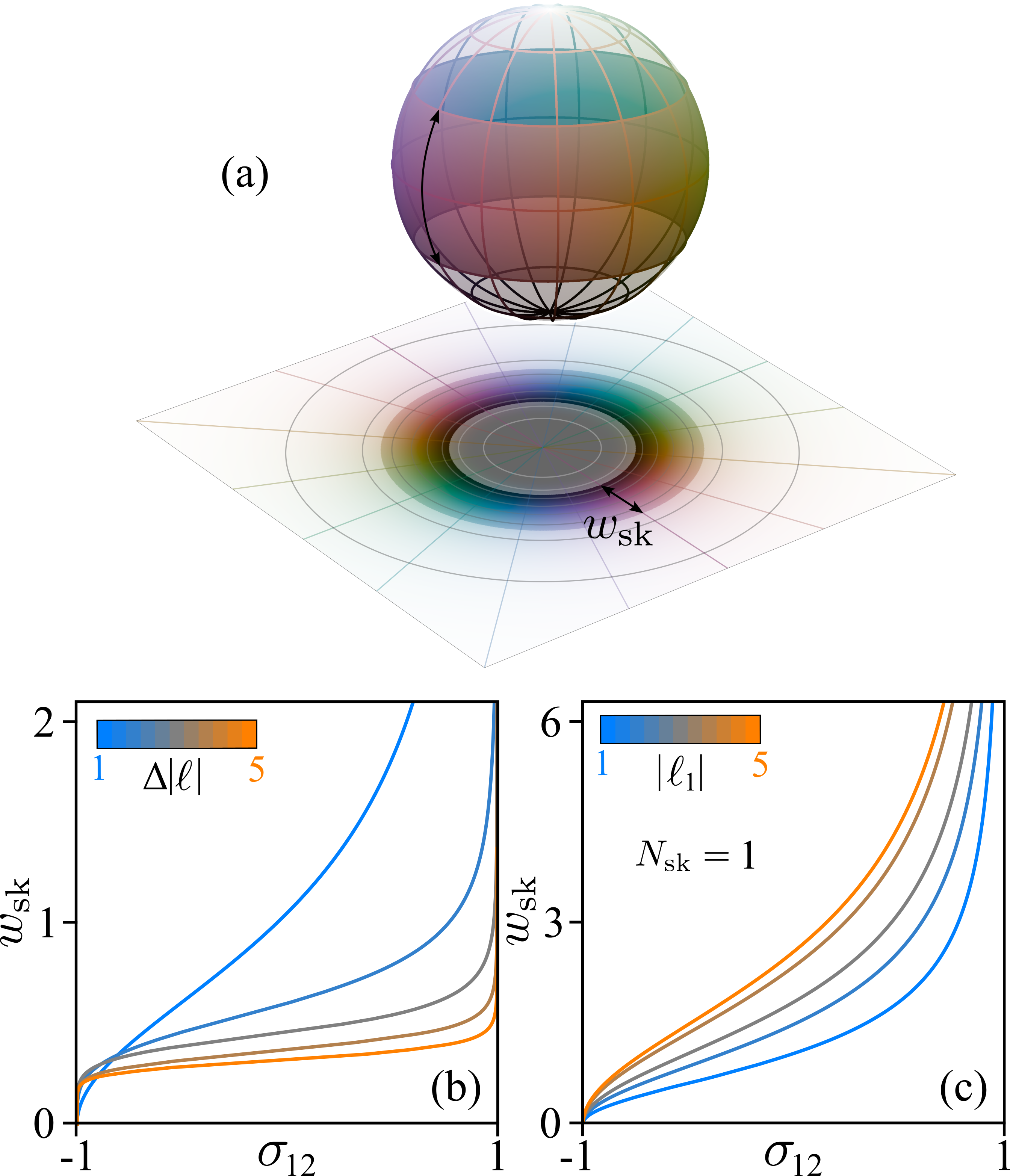}
    \caption{Skyrmion width $w_{\rm sk}$ highlighted in the transverse plane and its corresponding area covered in the PS (a). Skyrmion width as a function of the relative weight balance $\sigma_{12}$ for several topological charge difference $\Delta|\ell|=|\ell_2|-|\ell_1|$ (b) and the same for fixed $\Delta|\ell|=1$ and different values of $|\ell_1|$ (c).}
    \label{fig:SMF1}
\end{figure}

A skyrmionic polarization texture is constructed as the superposition of two vortices with different polarization states. The characteristic size of such a texture decreases as the energy is more distributed towards the vortex of higher order, and increases as the energy is more distributed towards the component of lower order. In a paraxial skyrmionic beam the polarization texture can be parametrized from the field ratio between two polarization components. Each component is a vortex of topological charges $\ell_1$ and $\ell_2$ (with $|\ell_2|>|\ell_1|$) so that the field ratio becomes
\begin{equation}\label{eq:ratioLG}
    \frac{\psi_1}{\psi_2}=\lr{\frac{r\sqrt{2}}{w(z)}}^{|\ell_1|-|\ell_2|}\sqrt{\frac{|\ell_2|!}{|\ell_1|!}\cdot\frac{1+\sigma_{12}}{1-\sigma_{12}}}e^{i\lr{\ell_1-\ell_1}\phi+i\Delta\mathcal{G}(z)},
\end{equation}
where $\Delta\mathcal{G}(z)=-\lr{|\ell_1|-|\ell_2|}\tan^{-1}\lr{z/z_R}$, $\sigma_{12}=\iint\lr{|\psi_1|^2-|\psi_2|^2}\rrm{d}x\rrm{d}y/\iint\lr{|\psi_1|^2+|\psi_2|^2}\rrm{d}x\rrm{d}y$ measures the relative weighting of both polarization components, $\sigma_{12}=1$ if the weight of $\psi_2$ is zero and thus all power is carried by the vortex of lower charge, $\sigma_{12}=-1$ if the weight of $\psi_1$ is zero and thus all power is carried by the vortex of higher charge, and $\sigma_{12}=0$ if both weights are equal. A characteristic size of the skyrmionic beam can be defined by evaluating the thickness of a disk $w_{\rm sk}$ in the transverse plane that covers half the surface of the Poincar\'e sphere, see Fig.~\ref{fig:SMF1}(a). 
We choose a spherical belt around the equator since for the case $|\ell_1|\neq0$ this provides a more appropriate measure of the area at which the polarization changes, as for large values of $|\ell_1|$ the polarization texture appears mainly around a ring centered away from the origin unlike for $|\ell_1|=0$, see for instance~\cite{Wang2025GenerationNumber}. Choosing the inner and outer radius of as $\rho^2\lr{r_{\rm in}}=\rho_{\rm in}^2=3$ and $\rho^2\lr{r_{\rm out}}=\rho_{\rm out}^2=1/3$ the ring covers exactly half the solid angle of the PS. From (\ref{eq:ratioLG}) we obtain the skyrmion width as
$$w_{\rm sk}(z)=\frac{w(z)}{\sqrt{2}}\lr{\frac{1+\sigma_{12}}{1-\sigma_{12}}\cdot\frac{|\ell_2|!}{|\ell_1|!}}^{\frac{1}{2}\frac{1}{|\ell_1|-|\ell_2|}}\times$$

\begin{equation}\label{eq:w_sk}
    \lr{\rho_{\rm in}^{\frac{1}{|\ell_1|-|\ell_2|}}-\rho_{\rm out}^{\frac{1}{|\ell_1|-|\ell_2|}}}.
\end{equation}
It is evident that Stokes skyrmions expand at the same rate as LG beams, regardless of the criterion to define their characteristic size, i.e., their size is proportional to the propagation-dependent beam size $w(z)$. In addition, the size of the texture $w_{\rm sk}$ can be shrunk down to arbitrarily small scales by distributing the energy towards the component of highest order $|\ell_2|$, regardless of the value of $|\ell_1|$ and $|\ell_2|$ as shown in~\ref{fig:SMF1} (b-c), from the $\lr{1+\sigma_{12}}$ factor in (\ref{eq:w_sk}) which can be arbitrarily close to zero. The property of polarization textures to cover large areas of the PS in a small region of the transverse plane (super-coverage) is a generalization of super-oscillations and happens at the expense of having low intensity of light~\cite{Maxwell2025StochasticFields}.

\subsection{Mode independence of polarization structure}\label{sec:A5}
Here we show that the non-transverse polarization texture around a vortex line does not depend on the shape of the vortex beam. We start by expanding an arbitrary vortex beam $\psi_\perp(r,\phi,z)$ with topological charge $\ell$ as a sum of LG modes, since these constitute a basis to express any paraxial field. Therefore we use Eq. (\ref{eq:LGbeam}) and write
\begin{equation}\label{eq:Et_mode_expansion}
    \psi_\perp(r,\phi,z)=\sum_{p=0}^\infty c_p \rrm{LG}_{p}^{\ell}(r,\phi,z)=\mathcal{S}(r,z)\mathcal{A}(r,z) e^{i\ell\phi},
\end{equation}
where $\mathcal{S}(r,z)$ accounts for the mode expansion
$$\mathcal{S}(r,z)=\sum_{p=0}^\infty c_pL_p^{|\ell|}\lr{\frac{2r^2}{w^2(z)}}\exp\lr{-i\mathcal{G}_p(z)},$$
and $\mathcal{A}(r,z)$ is a mode-independent amplitude factor
$$\mathcal{A}(r,z)=\frac{1}{w(z)}\lr{\frac{r\sqrt{2}}{w(z)}}^{|\ell|}\times$$
$$\times\exp\lrr{i\lr{\frac{kr^2}{2R(z)}+kz}-\frac{r^2}{w^2(z)}}.$$
The longitudinal field again can be calculated for $\ell>0$ and RCP and $\ell<0$ and LCP yielding
\begin{widetext}
\begin{equation}\label{eq:psi_z_expansion}
    \psi_z(r,\phi,z)=\frac{i\mathcal{A}(r,z)}{\sqrt{2}k}\lrr{S(r,z)\lr{\frac{2|\ell|}{r}-\frac{2r}{w^2(z)}+\frac{ikr}{2R(z)}}-\frac{4r}{w^2(z)}\sum_{p=0}^\infty c_{p+1}L_{p}^{|\ell|+1}\lr{\frac{2r^2}{w^2(z)}}}e^{\pm i\lr{|\ell|-1}\phi},
\end{equation} 
\end{widetext}
where upper and lower signs stand for $\ell>0$ and RCP and $\ell<0$ and LCP. Here we have used the following property of Laguerre polynomials
$$\dd{L_p^{|\ell|}\lr{\xi}}{\xi}=-L_{p-1}^{|\ell|+1}\lr{\xi}$$
for $p\geq1$, and zero otherwise. Now we take the asymptotic behavior for points close to the phase singularity $\lrr{r/w(z)\ll1}$, that is the lowest terms in $r$. The last sum $\sum_{p=0}^\infty c_{p+1}L_{p}^{|\ell|+1}\lr{2r^2/w^2(z)}$ is dominated by a constant term and so does the series expansion
$$\lim_{r/w(z)\to0}\mathcal{S}(r,z)=\sum_{p=0}^\infty c_{p}\frac{\Gamma\lr{|\ell|+p+1}}{\Gamma\lr{|\ell|+1}p!}.$$
Therefore, the leading term in Eq. (\ref{eq:psi_z_expansion}) behaves like $r^{-1}$, which is two orders of magnitude greater than the following terms (of the order of $r^1$), so that the longitudinal field around the vortex core reads
\begin{equation}\label{eq:Ez_mode_expansion_simplified}
    \psi_z(r,\phi,z)\approx\frac{i|\ell|\sqrt{2}}{kr}\mathcal{A}\lr{r,z}\mathcal{S}\lr{r,z}e^{\pm i\lr{|\ell|-1}\phi},
\end{equation}
where again upper and lower signs refer to RCP and $\ell>0$, LCP and $\ell<0$, the longitudinal field being negligible with respect to the transverse field otherwise. With an arbitrary transverse polarization state given by (\ref{eq:u_perp}) we perform again the ratio between longitudinal (\ref{eq:Ez_mode_expansion_simplified}) and transverse (\ref{eq:Et_mode_expansion}) fields and we obtain that the polarization texture is governed by the complex field
$$\rho e^{i\gamma}=\frac{\psi_z}{\psi_\perp}=\frac{i|\ell|}{kr}\sqrt{1-\rrm{sgn}\lr{\ell}\sigma_z}e^{-i\rrm{sgn}\lr{\ell}\lr{\phi+\phps/2}},$$
which is indeed independent of the expansion coefficients, in other words the shape of the vortex beam, and also independent of the propagation distance, i.e. propagation-invariant. 

Finally, we could also consider the anisotropy of the vortex beam, which accounts for beam imperfections in real scenarios. Letting $\ell>0$, the anisotropy of the vortex phase in Eq. (\ref{eq:Et_mode_expansion}) is accounted by introducing a component with opposite topological charge $-\ell$, such as
\begin{equation}\label{eq:Et_anisotropic}
    \psi_\perp(r,\phi,z)=\mathcal{S}(r,z)\mathcal{A}(r,z)\lrr{c_{+}e^{i\ell\phi}+c_ {-}e^{-i\ell\phi}},
\end{equation}
where the components with positive (negative) topological charges are weighted by the complex weights $c_{+}$ $\lr{c_{-}}$, with $|c_{+}|^2+|c_{-}|^2=1$. Again with arbitrary polarization state (\ref{eq:u_perp}), the longitudinal field only arises from the component with RCP and $\ell>0$ and with LCP and $-\ell<0$, which for regions around the beam axis $r\ll w(z)$ yields
$$\psi_z(r,\phi,z)\approx\frac{i|\ell|\sqrt{2}}{kr}\mathcal{A}\lr{r,z}\mathcal{S}\lr{r,z}\times $$
\begin{equation}\label{eq:Ez_anisotropic}
\times\lrr{\tilde{c}_{+}e^{i\lr{\ell-1}\phi}+\tilde{c}_{-}e^{i\lr{-\ell+1}\phi}}
\end{equation}
with $\tilde{c}_{+}=c_{+}\cos\lr{\apps}\exp\lr{-i\phps/2}$ and $\tilde{c}_{-}=c_{-}\sin\lr{\apps}\exp\lr{i\phps/2}$ being complex coefficients. Finally, with (\ref{eq:Et_anisotropic}) and (\ref{eq:Ez_anisotropic}) the field ratio for a generic anisotropic vortex line reads
\begin{equation}\label{eq:ratio_anisotropic}
    \rho e^{i\gamma}=\frac{\psi_z}{\psi_\perp}=\frac{i|\ell|\sqrt{2}}{kr}\frac{\tilde{c}_{+}e^{i\lr{\ell-1}\phi}+\tilde{c}_{-}e^{i\lr{-\ell+1}\phi}}{c_{+}e^{i\ell\phi}+c_ {-}e^{-i\ell\phi}},
\end{equation}
which presents the same radial behavior $\lr{\sim 1/r}$ as in an isotropic vortex (\ref{eq:ratio}) but with a different azimuthal distribution. Yet, the propagation distance still does not appear in the polarization structure (\ref{eq:ratio_anisotropic}). This confirms that after introducing a perturbation in the field which distorts the vortex structure, the polarization texture will still maintain a non-spreading propagation in free-space.
%\clearpage

\bibliography{references,notes}
\end{document}